\documentclass[aps,pra,showpacs,twocolumn]{revtex4-1}
\usepackage{amssymb}
\usepackage{amsmath}
\usepackage{graphicx}
\usepackage{epsfig}

\begin{document}

\title{Resonant generation of $p$-wave Cooper pair in non-Hermitian Kitaev
chain at exceptional point}
\author{X. M. Yang and Z. Song}
\email{songtc@nankai.edu.cn}
\affiliation{School of Physics, Nankai University, Tianjin 300071,
China}

\begin{abstract}
We investigate a non-Hermitian extension of Kitaev chain by considering
imaginary $p$-wave pairing amplitudes. The exact solution shows that the
phase diagram consists two phases with real and complex \ Bogoliubov-de-gens
spectra, associated with $\mathcal{PT}$-symmetry breaking, which is
separated by a hyperbolic exceptional line. The exceptional points (EPs)
correspond to a specific Cooper pair state $( 1+c_{k}^{\dagger
}c_{-k}^{\dagger }) \left\vert 0\right\rangle $\ with movable $k$ when the
parameters vary along the exceptional line. The non-Hermiticity around EP
supports resonant generation of such a pair state from the vacuum state $%
\left\vert 0\right\rangle $ of fermions via the critical dynamic process. In
addition, we propose a scheme to generate a superconducting state through a
dynamic method.
\end{abstract}

\maketitle

\section{Introduction}

The Kitaev model is a lattice model of a $p$-wave superconducting wire,
which realize Majorana zero modes at the ends of the chain \cite{Kitaev}.
This has been demonstrated by unpaired Majorana modes exponentially
localized at the ends of open Kitaev chains \cite{Sarma,Stern,Alicea}. The
main feature of this model originates from the pairing term, which violates
the conservation of the fermion number but preserves its parity, leading to
the superconducting phase. The amplitudes for pair creation and annihilation
plays an important role in the existence of the gapped superconducting
phase. In general, most of the investigations on this model has focused on
the case with a Hermitian pairing term. A non-Hermitian term is no longer
forbidden both in theory and experiment since the discovery that a certain
class of non-Hermitian Hamiltonians could exhibit entirely real spectra \cite%
{Bender,Bender1}. The origin of the reality of the spectrum of a
non-Hermitian Hamiltonian is the pseudo-Hermiticity of the Hamiltonian
operator \cite{Ali1}. It motives a non-Hermitian extension of the Kitaev
model. Many contributions have been devoted to non-Hermitian Kitaev models
\cite{Law,Tong,Yuce,You,Klett,Menke} and Ising models \cite{ZXZ,LCPRB}
within the pseudo-Hermitian framework. Also, the experimental schemes for
realizing the Kitaev model and related non-Hermitian systems has been
presented in Refs. \cite{Franz} and \cite{Ueda}, respectively. In addition,
the peculiar features of a non-Hermitian system do not only manifest in
statics but also dynamics. From the perspective of non-Hermitian quantum
mechanics, it is also a challenge to deal with many-particle dynamics.

In this paper, we investigate a non-Hermitian extension of Kitaev chain by
considering imaginary $p$-wave pairing amplitudes. Theoretically, an open
system is regarded as a subsystem of an infinite Hermitian system, while a
non-Hermitian Hamiltonian is introduced to describe the physics of the
subsystem in a phenomenological way \cite{Rotter}. Non-Hermitian $p$-wave
pairing amplitudes may arise from the case, in which subsystem and the
surrounding system are in the superconducting phase.\textbf{\ }When the
whole system is in some non-equilibrium superconducting\textbf{\ }states%
\textbf{, }the subsystem should be effectively described by a non-Hermitian
pair creation and annihilation.\textbf{\ }As a concrete step toward this,
the quantum tunneling of particle pairs has been studied for two weakly
interacting systems as a superconducting tunnel junction \cite{Matthews}.%
\textbf{\ }Non-Hermitian systems exhibit many peculiar dynamic behaviors
that never occurred in Hermitian systems. One of the remarkable features is
the dynamics at exceptional point (EP) \cite{HeissEP,RotterEPJPA,HXu} or
spectral singularity (SS) \cite{AMprl,SLHprb,AAA,BFS,ZXZ2}, where the system
has a coalescence state. In this work, we focus on the EP-related dynamic
behavior for the many-body system.

Based on the exact solution, we find that the exceptional line is a
hyperbolic in parameter space, which separates two regions with real and
complex Bogoliubov-de-gens spectra, associated with unbroken and broken $%
\mathcal{PT}$-symmetric phase, respectively. The EPs move in $k$ space when
the parameters vary along the exceptional line. In addition, the critical
dynamics supports resonant generation of $p$-wave Cooper pair: a specific
pair state $c_{k_{\mathrm{c}}}^{\dagger }c_{-k_{\mathrm{c}}}^{\dagger
}\left\vert 0\right\rangle $\ with selecting opposite momentum $k_{\mathrm{c}%
}$\ can be generated from the vacuum state $\left\vert 0\right\rangle $\ of
fermions by natural time evolution. The selected $k_{\mathrm{c}}$\ ranges
over the Brillouin zone, determined by the parameters. The underlying
mechanism stems from the critical dynamics around the EP, that projects an
initial state on the coalescing state. Our work also exemplifies the dynamic
nature of a non-Hermitian interacting many-particle system. As an
application, it provides alternative way to generate a superconducting state
from an empty state via critical dynamic process rather than cooling down
the temperature.

This paper is organized as follows. In Section \ref{Non-Hermitian Kitaev
model}, we describe the model Hamiltonian. In Section \ref{Phase diagram},
based on the solutions, we present the phase diagram. In Section \ref%
{Dynamics}, we study the dynamics in the unbroken-symmetry region, including
the time evolution at exceptional line. In Section \ref{p-wave pair
generation}, we focus on the critical dynamics for vacuum state as initial
state. In Section \ref{superconducting state}, we propose a scheme to
generate a superconducting state.\ Finally, we give a summary and discussion
in Section \ref{sec_summary}.

\section{Non-Hermitian Kitaev model}

\label{Non-Hermitian Kitaev model} We consider the following fermionic
Hamiltonian on a lattice of length $N$

\begin{eqnarray}
\mathcal{H} &=&\sum\limits_{j=1}^{N}[-Jc_{j}^{\dag }c_{j+1}+\mathrm{H.c.}%
-i\Delta c_{j}^{\dag }c_{j+1}^{\dag }  \label{NH} \\
&&-i\Delta c_{j+1}c_{j}+\mu \left( 2n_{j}-1\right) ],  \notag
\end{eqnarray}%
where $c_{j}^{\dag }$ $(c_{j})$\ is a fermionic creation (annihilation)
operator on site $j$, $n_{j}=c_{j}^{\dag }c_{j}$, $J$ the tunneling rate, $%
\mu
$ the chemical potential, and $i\Delta $\ the strength of the $p$-wave pair
creation (annihilation). We define $c_{N+1}=c_{1}$ for periodic boundary
condition. The Hamiltonian (\ref{NH}) is known to have a rich phase diagram
in its Hermitian version, i.e., $i\Delta \rightarrow \Delta $, which is a
spin-polarized $p$-wave superconductor in one dimension. This system is
known to have topological phases in which there is a zero energy Majorana
mode at each end of a long chain. It is also the fermionized version of the
familiar one-dimensional transverse-field Ising model \cite{Pfeuty}, which
is one of the simplest solvable models exhibiting quantum criticality and
demonstrating a quantum phase transition with spontaneous symmetry breaking
\cite{SachdevBook}. In this work, we consider a non-Hermitian extension by
imaginary pairing amplitude $i\Delta $. Comparing with the non-Hermitian
Kitaev model in previous works \cite{Law,Tong,Yuce,You,Klett,Menke,LCPRB},
the present model has parity-time-reversal ($\mathcal{PT}$) symmetry (proved
below) and its non-Hermiticity arises from the imaginary pairing term rather
than from the on-site potential term. We will show that the quasi-particle
spectrum can have two movable EPs, resulting in some exclusive features
different from its Hermitian version.

Before solving the Hamiltonian, it is profitable to investigate the symmetry
of the system. By the direct derivation, we have $\left[ \mathcal{PT},%
\mathcal{H}\right] =0$, where the antilinear time reversal operator $%
\mathcal{T}$ has the function $\mathcal{T}i\mathcal{T=-}i$, and $\left(
\mathcal{P}\right) ^{-1}c_{l}\mathcal{P}=c_{N-l+1}$. As a usual
pseudo-Hermitian system \cite{Ali}, the $\mathcal{PT}$\ symmetry in the
present model plays the same role to the phase diagram. The spectrum of $%
\mathcal{H}$\ can be real if all the eigenstates can be written as a $%
\mathcal{PT}$-symmetric form, while complex when the corresponding
eigenstates break the $\mathcal{PT}$-symmetry.\ The concept of EPs in this
paper specifies the locations in the parameter space, at which the complex
spectrum starts to appear (In general, an EP is any point with coalescing
state). We concentrate our work on the real-spectrum (or unbroken symmetry)
region, avoiding the exponentially increased Dirac probability.\

\begin{figure*}[tbp]
\includegraphics[ bb=0 0 1102 362, width=0.9\textwidth,
clip]{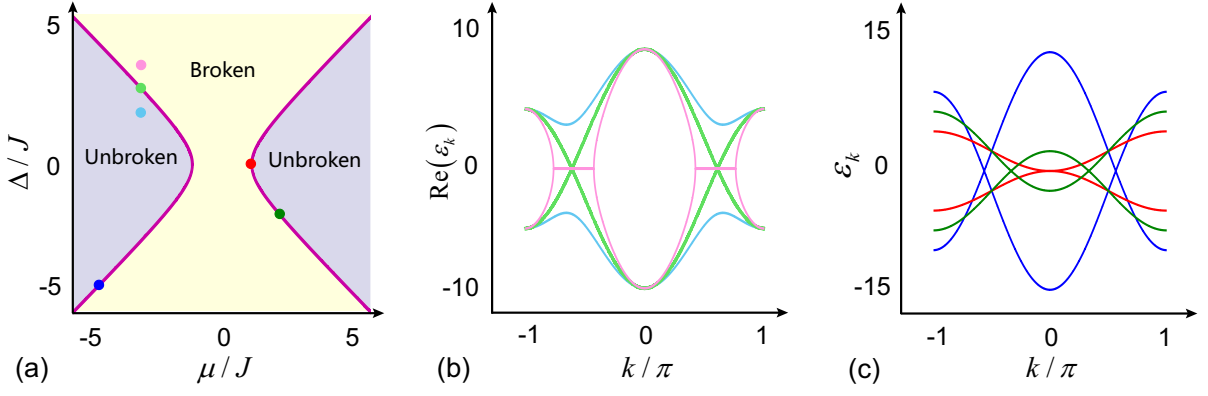}
\caption{(a) The phase diagram of the non-Hermitian Kitaev model with
imaginary p-wave pairing amplitudes. The phase boundary is a hyperbolic
exceptional lines (darkmagenta), which separate two regions with real
(purple) and complex (yellow) Bogoliubov-de-gens spectra, associated with
unbroken $\mathcal{PT}$-symmetric phase and broken $\mathcal{PT}$-symmetric
phase respectively. (b) Real part of quasi-particle spectra for three
typical points indicated in (a), representing unbroken phase (lightblue), EP
line (lightgreen) and broken phase (lightpink), respectively. (c)
Quasi-particle spectra for three typical points at the phase boundary line
indicated in (a). The corresponding EPs are movable and merges at fixed $k=0$
or $k=\protect\pi$.}
\label{fig1}
\end{figure*}

In this work, we focus on the dynamics of such a superconducting system,
which motivates a more systematic study. Taking the Fourier transformation

\begin{equation}
c_{j}=\frac{1}{\sqrt{N}}\sum\limits_{k}e^{ikj}c_{k},
\end{equation}%
for the Hamiltonian (\ref{NH}), with wave vector $k\in (-\pi ,\pi ]$, we
have
\begin{eqnarray}
\mathcal{H} &=&-\sum_{k}[2\left( J\cos k-\mu \right) c_{k}^{\dag }c_{k}
\notag \\
&&+\Delta \sin k(c_{-k}c_{k}+c_{-k}^{\dag }c_{k}^{\dag })+\mu ].
\end{eqnarray}%
For the convenience of further analysis, we express the Hamiltonian by using
the Nambu representation
\begin{eqnarray}
\mathcal{H} &=&\sum_{\pi >k>0}\mathcal{H}_{k}, \\
\mathcal{H}_{k} &=&2\left(
\begin{array}{cc}
c_{k}^{\dag } & c_{-k}%
\end{array}%
\right) \left(
\begin{array}{cc}
\mu -J\cos k & \Delta \sin k \\
-\Delta \sin k & J\cos k-\mu%
\end{array}%
\right) \left(
\begin{array}{c}
c_{k} \\
c_{-k}^{\dag }%
\end{array}%
\right) ,  \notag \\
&&
\end{eqnarray}%
where the Hamiltionian $\mathcal{H}_{k}$\ in each invariant subspace
satisfies the commutation relation%
\begin{equation}
\left[ \mathcal{H}_{k},\mathcal{H}_{k^{\prime }}\right] =0.
\end{equation}

This allows us to treat the diagonalization and the dynamics governed by $%
\mathcal{H}_{k}$\ individually. So far the procedure is the same as those
for solving the Hermitian version of $\mathcal{H}$. To diagonalize a
non-Hermitian Hamiltonian, we should introduce the Bogoliubov transformation
in the complex version:%
\begin{equation}
\left\{
\begin{array}{l}
\gamma _{k}=\cos \theta _{k}c_{k}-i\sin \theta _{k}c_{-k}^{\dag }, \\
\overline{\gamma }_{k}=\cos \theta _{k}c_{k}^{\dagger }+i\sin \theta
_{k}c_{-k},%
\end{array}%
\right.
\end{equation}%
where the complex angle $\theta _{k}$\ is determined by%
\begin{equation}
\tan \left( 2\theta _{k}\right) =\frac{i\Delta \sin k}{\mu -J\cos k}.
\end{equation}%
It is a crucial step to diagonalize a non-Hermitian Hamiltonian, which
essentially establishes the biorthogonal modes. It is easy to check that the
complex Bogoliubov modes $\left( \gamma _{k},\overline{\gamma }_{k}\right) $
satisfy the anticommutation relations%
\begin{eqnarray}
\left\{ \gamma _{k},\overline{\gamma }_{k^{\prime }}\right\} &=&\delta
_{k,k^{\prime }}, \\
\left\{ \gamma _{k},\gamma _{k^{\prime }}\right\} &=&\left\{ \overline{%
\gamma }_{k},\overline{\gamma }_{k^{\prime }}\right\} =0,  \notag
\end{eqnarray}%
which result in the diagonal form of the Hamiltonian%
\begin{equation}
\mathcal{H}=\sum\limits_{k}\varepsilon _{k}\left( \overline{\gamma }%
_{k}\gamma _{k}-\frac{1}{2}\right) .  \label{diagonal H}
\end{equation}%
Here
\begin{equation}
\varepsilon _{k}=2\sqrt{\left( \mu -J\cos k\right) ^{2}-\Delta ^{2}\sin ^{2}k%
}.
\end{equation}%
is the dispersion relation of the quasiparticle\textbf{. }Note that the
Hamiltonian $\mathcal{H}$\ is still non-Hermitian due to the fact that $%
\overline{\gamma }_{k}\neq \gamma _{k}^{\dag }$. In addition, quasi-spectrum
$\varepsilon _{k}$\ can be real or imaginary, but not zero since the complex
Bogoliubov modes $\left( \gamma _{k},\overline{\gamma }_{k}\right) $ is not
well-defined if $\varepsilon _{k}=0$, which will be discussed in next
section.

\section{Phase diagram}

\label{Phase diagram} According to the theory for a pseudo-Hermitian system
\cite{Ali}, the whole parameter space consists of two kinds of regions,
symmetry-unbroken one with a fully real spectrum and symmetry-broken one
with a complex spectrum, which originates from the over-threshold imaginary
pairing amplitudes. The reason can be seen from the following derivation.
For a given $k$, the Hamiltonian $\mathcal{H}_{k}$\ in the basis ($%
\left\vert 0\right\rangle _{k}\left\vert 0\right\rangle _{-k}$, $\left\vert
1\right\rangle _{k}\left\vert 1\right\rangle _{-k}$, $\left\vert
1\right\rangle _{k}\left\vert 0\right\rangle _{-k}$, $\left\vert
0\right\rangle _{k}\left\vert 1\right\rangle _{-k}$) is expressed as $%
4\times 4$ matrix%
\begin{equation}
\mathcal{H}_{k}=2\left(
\begin{array}{cccc}
J\cos k-\mu  & -\Delta \sin k & 0 & 0 \\
\Delta \sin k & \mu -J\cos k & 0 & 0 \\
0 & 0 & 0 & 0 \\
0 & 0 & 0 & 0%
\end{array}%
\right) .
\end{equation}%
$\allowbreak $The eigenstates $\left\vert \psi _{k\mathrm{\lambda }}^{\pm
}\right\rangle $\ ($\mathrm{\lambda }=\mathrm{e,o}$ denotes the even/odd
parity of the particle number) are
\begin{eqnarray}
\left\vert \psi _{k\mathrm{e}}^{\pm }\right\rangle  &=&\frac{1}{\sqrt{\Omega
_{\mathrm{nh}}^{\pm }}}\left( \left\vert 0\right\rangle _{k}\left\vert
0\right\rangle _{-k}+\beta _{k}^{\pm }\left\vert 1\right\rangle
_{k}\left\vert 1\right\rangle _{-k}\right) , \\
\left\vert \psi _{k\mathrm{o}}^{+}\right\rangle  &=&\left\vert
1\right\rangle _{k}\left\vert 0\right\rangle _{-k},\left\vert \psi _{k%
\mathrm{o}}^{-}\right\rangle =\left\vert 0\right\rangle _{k}\left\vert
1\right\rangle _{-k},
\end{eqnarray}%
where $\Omega _{\mathrm{nh}}^{\pm }=1+\left\vert \beta _{k}^{\pm
}\right\vert ^{2}$ is the normalization coefficient in the context of Dirac
inner product with%
\begin{equation}
\beta _{k}^{\pm }=\frac{\Delta \sin k}{J\cos k-\mu \pm \varepsilon _{k}/2},
\end{equation}%
and corresponding energies are%
\begin{equation}
\varepsilon _{k\mathrm{e}}^{\pm }=\pm \varepsilon _{k},\varepsilon _{k%
\mathrm{o}}^{\pm }=0.
\end{equation}

We note that%
\begin{equation}
\left\{
\begin{array}{cc}
\mathcal{PT}\left\vert \psi _{k\mathrm{e}}^{\pm }\right\rangle =\left\vert
\psi _{k\mathrm{e}}^{\pm }\right\rangle , & \text{for }\left( \varepsilon
_{k}\right) ^{2}>0 \\
\mathcal{PT}\left\vert \psi _{k\mathrm{e}}^{\pm }\right\rangle =\left\vert
\psi _{k\mathrm{e}}^{\mp }\right\rangle , & \text{for }\left( \varepsilon
_{k}\right) ^{2}<0%
\end{array}%
\right. ,
\end{equation}%
while
\begin{equation}
\mathcal{PT}\left\vert \psi _{k\mathrm{o}}^{\pm }\right\rangle =e^{\mp
ik}\left\vert \psi _{k\mathrm{o}}^{\pm }\right\rangle ,
\end{equation}%
for both unbroken and broken $\mathcal{PT}$-symmetric phases, where we used
the relation

\begin{equation}
\left( \mathcal{PT}\right) ^{-1}c_{k}^{\dagger }\mathcal{PT}%
=e^{-ik}c_{k}^{\dagger }.
\end{equation}%
As expected, the symmetry of the eigenstates is associated with the reality
of the energy level. An eigenstate of $\mathcal{H}$ is constructed as the
form%
\begin{equation}
\left\vert \Psi \right\rangle =\prod_{\pi >k>0}\left\vert \varphi
_{k}^{\lambda }\right\rangle ,
\end{equation}%
where the index $\lambda =1,2,3,4$ labels the eigenstate in each $k$ sector,
$\left\vert \varphi _{k}^{1,2}\right\rangle =\left\vert \psi _{k\mathrm{e}%
}^{+.-}\right\rangle $ and $\left\vert \varphi _{k}^{3,4}\right\rangle
=\left\vert \psi _{k\mathrm{o}}^{+.-}\right\rangle $, with the eigen energy%
\begin{equation}
E=\sum_{\pi >k>0}\epsilon _{k}^{\lambda },
\end{equation}%
with $\epsilon _{k}^{1,2}=\varepsilon _{k\mathrm{e}}^{+,-}$ and $\epsilon
_{k}^{3,4}=0$. Therefore, the reality of $\varepsilon _{k}$\ determines the
reality of the spectrum of $\mathcal{H}$, since a single imaginary $%
\varepsilon _{k}$\ can result in the complex spectrum of $\mathcal{H}$. A
quantum phase transition occurs when the complex spectrum appears. Then the
phase boundary of $\mathcal{H}$\ locates at the touching point of curve $%
\varepsilon _{k}$\ at $k$ axis. The phase boundary (or EP line) in parameter
space ($\mu -\Delta $ plane) is determined by the equations \cite{JLPRA}.%
\begin{equation}
\varepsilon _{k_{\mathrm{c}}}=\left[ \frac{\partial \varepsilon _{k}}{%
\partial k}\right] _{k=k_{\mathrm{c}}}=0.
\end{equation}%
The boundary is obtained as
\begin{equation}
\mu _{\mathrm{c}}^{2}-\Delta _{\mathrm{c}}^{2}=J^{2}  \label{boundary}
\end{equation}%
with
\begin{equation}
k_{\mathrm{c}}=\arccos \frac{J}{\mu _{\mathrm{c}}}.
\end{equation}%
In this situation, $\mathcal{H}_{k_{\mathrm{c}}}$\ cannot be expressed as
the complex Bogoliubov modes $\left( \gamma _{k_{\mathrm{c}}},\overline{%
\gamma }_{k_{\mathrm{c}}}\right) $ since the matrix of $\mathcal{H}_{k_{%
\mathrm{c}}}$\ in even particle number sector has a Jordan block form%
\begin{equation}
M_{\mathrm{c}}=\frac{-2\Delta _{\mathrm{c}}^{2}}{\mu _{\mathrm{c}}}\left(
\begin{array}{cc}
1 & -1 \\
1 & -1%
\end{array}%
\right) ,
\end{equation}%
which hence is no longer diagonalizable. Two eigenvectors of $M_{\mathrm{c}}$%
\ coalesce to a single one $\left( 1,1\right) ^{\mathrm{T}}$, leading to a
set of coalescing eigenstates of $\mathcal{H}$, including the coalescing\
groundstate. Remarkably, $\mathcal{H}_{k_{\mathrm{c}}}$\ governs a peculiar
dynamics, which is the focus of this work. The phase diagram on parameter $\mu -\Delta $\ plane is plotted in Fig. \ref{fig1}(a). The real part of
quasi-particle spectra for several typical points in symmetry-unbroken,
broken phase phases, and on EP line are plotted in Fig. \ref{fig1}(b) and
(c). It shows that the pair of EPs are movable and meet at a fixed point.
Such a gapless phase is different from its Hermitian version, where the band
touching point is degenerate point. It will result in different dynamical
behavior in the non-Hermitian Kitaev model, especially near the phase
boundary.

\begin{figure*}[tbp]
\includegraphics[ bb=4 75 467 255, width=0.59\textwidth, clip]{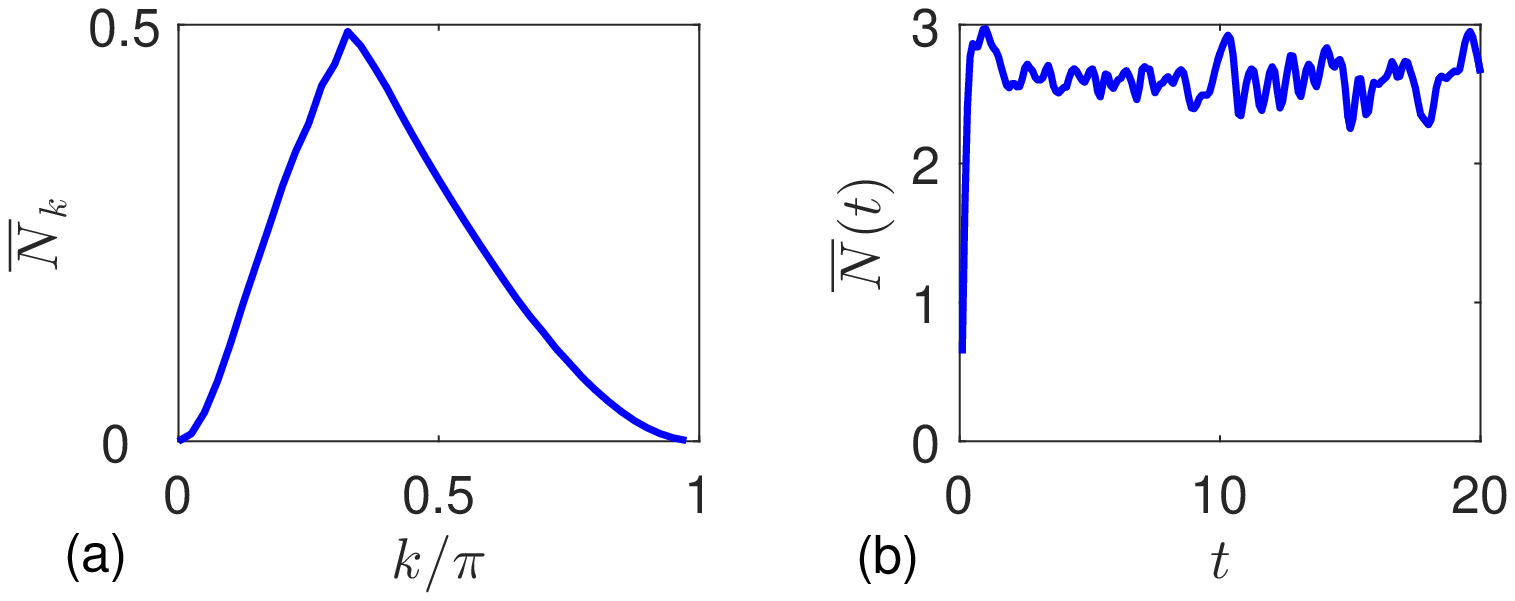} %
\includegraphics[ bb=460 75 717 255, width=0.335\textwidth,clip]{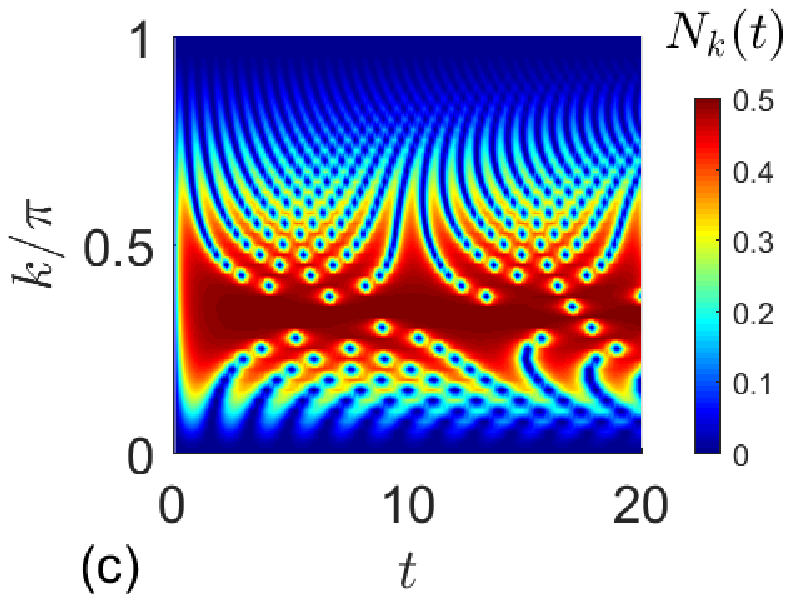}
\caption{Plots of (a) $\overline{N}_{k}$, (b) ${\overline{N}}(t)$, and (c) $%
N_{k}(t)$, which is defined in Eqs. (\protect\ref{aNk}), (\protect\ref{aNt})
and (\protect\ref{Nkt}). The parameters are $N=40$ and $(J,\Delta ,\protect%
\mu )=(1,\protect\sqrt{3},2)$.}
\label{fig2}
\end{figure*}

\section{Dynamics}

\label{Dynamics} We study the dynamics in the unbroken-symmetry region, in
which $\varepsilon _{k}$\ is always real, including the time evolution at
exceptional line. Based on the above analysis, the dynamics is governed by
the time evolution operator%
\begin{equation}
U(t)=\exp (-i\mathcal{H}t)=\prod_{\pi >k>0}U_{k}(t),
\end{equation}%
where%
\begin{equation}
U_{k}(t)=\exp (-i\mathcal{H}_{k}t).
\end{equation}%
The explicit form of $U_{k}(t)$\ is determined by the diagonal form of $%
\mathcal{H}_{k}$, i.e.,%
\begin{equation}
\mathcal{H}_{k}=\varepsilon _{k}\left( \overline{\gamma }_{k}\gamma _{k}+%
\overline{\gamma }_{-k}\gamma _{-k}-1\right)
\end{equation}%
However, one of an exclusive features of a non-Hermitian system is that $%
\mathcal{H}_{k}$\ is undiagonalizable when $k=k_{\mathrm{c}}$.\ Therefore,
we will deal with $U_{k}(t)$\ in two aspects.

(i) In the case of $k\neq k_{\mathrm{c}}$, we have%
\begin{eqnarray}
U_{k}(t) &=&2\left[ \cos \left( \varepsilon _{k}t\right) -1\right] \bar{%
\gamma}_{k}\gamma _{k}\bar{\gamma}_{-k}\gamma _{-k} \\
&&+\left( 1-e^{i\varepsilon _{k}t}\right) \left( \bar{\gamma}_{-k}\gamma
_{-k}+\bar{\gamma}_{k}\gamma _{k}\right) +e^{i\varepsilon _{k}t},  \notag
\end{eqnarray}%
where we have used the identity $\left( \overline{\gamma }_{k}\gamma
_{k}\right) ^{2}=\overline{\gamma }_{k}\gamma _{k}$. This result is also
valid for imaginary $\varepsilon _{k}$. The vacuum state of $\gamma _{k}$\
is constructed as $\left\vert \mathrm{Vac}\right\rangle _{k}=\gamma
_{k}\left\vert 0\right\rangle $, where $\left\vert 0\right\rangle $\ is the
vacuum state of $c_{k}$. Four states ($\left\vert 00\right\rangle _{k}$, $%
\left\vert 11\right\rangle _{k}$, $\left\vert 10\right\rangle _{k}$, $%
\left\vert 01\right\rangle _{k}$)=($\left\vert \mathrm{Vac}\right\rangle
_{k}\left\vert \mathrm{Vac}\right\rangle _{-k}$, $\bar{\gamma}_{k}\bar{\gamma%
}_{-k}\left\vert \mathrm{Vac}\right\rangle _{k}\left\vert \mathrm{Vac}%
\right\rangle _{-k}$, $\bar{\gamma}_{k}\left\vert \mathrm{Vac}\right\rangle
_{k}\left\vert \mathrm{Vac}\right\rangle _{-k}$, $\bar{\gamma}%
_{-k}\left\vert \mathrm{Vac}\right\rangle _{k}\left\vert \mathrm{Vac}%
\right\rangle _{-k}$)\ are both the eigenstates of $\mathcal{H}_{k}$. The
time evolution of such four states are%
\begin{equation}
U_{k}(t)\left(
\begin{array}{c}
\left\vert 00\right\rangle _{k} \\
\left\vert 11\right\rangle _{k} \\
\left\vert 10\right\rangle _{k} \\
\left\vert 01\right\rangle _{k}%
\end{array}%
\right) =\left(
\begin{array}{c}
\exp \left( i\varepsilon _{k}t\right) \left\vert 00\right\rangle _{k} \\
\exp \left( -i\varepsilon _{k}t\right) \left\vert 11\right\rangle _{k} \\
\left\vert 10\right\rangle _{k} \\
\left\vert 01\right\rangle _{k}%
\end{array}%
\right) ,
\end{equation}%
which indicates that it looks like the one in Hermitian system if $%
\varepsilon _{k}$\ is real. The corresponding Dirac probability, $\left\vert
U_{k}(t)\left\vert mn\right\rangle _{k}\right\vert ^{2}$\ ($m,n=1,0$) is
conservative. However, the Dirac probability of a superposition of such two
eigenstates in even particle number subspace is periodic function of time
with period $\pi /\varepsilon _{k}$. It is noted that when $k$\ tends to $k_{%
\mathrm{c}}$, this period goes to infinite (or non-period), which is one of
properties of the critical dynamics.

(ii) In the case of $k=k_{\mathrm{c}}$, $\mathcal{H}_{k_{\mathrm{c}}}$
cannot be expressed as the complex Bogoliubov modes $\left( \gamma _{k_{%
\mathrm{c}}},\overline{\gamma }_{k_{\mathrm{c}}}\right) $. Nevertheless, we
can rewrite $\mathcal{H}_{k_{\mathrm{c}}}$\ in the form
\begin{equation}
\mathcal{H}_{k_{\mathrm{c}}}=-\frac{2\Delta _{\mathrm{c}}^{2}}{\mu _{\mathrm{%
c}}}\left( s_{k_{\mathrm{c}}}^{z}+is_{k_{\mathrm{c}}}^{y}\right) ,
\end{equation}%
by introducing pseudo-spin operators \cite{ZGPRL}%
\begin{eqnarray}
s_{k}^{x} &=&\frac{1}{2}\left( c_{-k}^{\dag }c_{k}^{\dag
}+c_{k}c_{-k}\right) ,  \notag \\
s_{k}^{y} &=&\frac{1}{2i}\left( c_{-k}^{\dag }c_{k}^{\dag
}-c_{k}c_{-k}\right) ,  \notag \\
s_{k}^{z} &=&\frac{1}{2}\left( c_{k}^{\dag }c_{k}+c_{-k}^{\dag
}c_{-k}-1\right) ,
\end{eqnarray}%
which satisfy Lie algebra $[s_{k}^{\alpha },s_{k}^{\beta }]=i\varepsilon
_{\alpha \beta \gamma }s_{k}^{\gamma }$, with $\varepsilon _{\alpha \beta
\gamma }$ being the Levi-Civita symbol. The corresponding time evolution
operator has the form
\begin{equation}
U_{k_{\mathrm{c}}}(t)=\exp (-i\mathcal{H}_{k_{\mathrm{c}}}t)=1-i\mathcal{H}%
_{k_{\mathrm{c}}}t,
\end{equation}%
based on the identity $\left( s_{k}^{z}+is_{k}^{y}\right) ^{2}=0$, or $%
\left( \mathcal{H}_{k_{\mathrm{c}}}\right) ^{2}=0$.

Obviously, the coalescing eigenstate of $\mathcal{H}_{k_{\mathrm{c}}}$\ is
the spin-up state in $x$ direction, $s_{k_{\mathrm{c}}}^{x}\left\vert
x\right\rangle =\frac{1}{2}\left\vert x\right\rangle $, and the
corresponding eigenstates are%
\begin{equation}
\left\vert \pm x\right\rangle =\left\vert 0\right\rangle _{k}\left\vert
0\right\rangle _{-k}\pm \left\vert 1\right\rangle _{k}\left\vert
1\right\rangle _{-k}.
\end{equation}%
Then the dynamics of the Jordan block is very clear, i.e.,%
\begin{eqnarray}
U_{k_{\mathrm{c}}}(t)\left\vert x\right\rangle &=&\left\vert x\right\rangle ,
\\
U_{k_{\mathrm{c}}}(t)\left\vert -x\right\rangle &=&\left\vert
-x\right\rangle +i4\left( \Delta _{\mathrm{c}}^{2}/\mu _{\mathrm{c}}\right)
t\left\vert x\right\rangle .
\end{eqnarray}%
Any initial state with component $\left\vert x\right\rangle $ obeys a
non-periodic (or infinite period) dynamics, which accords with the dynamics
of $\mathcal{H}_{k}$\ with $k\rightarrow k_{\mathrm{c}}$. In addition, the
evolved state $U_{k_{\mathrm{c}}}(t)\left\vert -x\right\rangle $ converges
to $\left\vert x\right\rangle $\ as time increases. This property also
appears in the dynamics of $\mathcal{H}_{k}$\ with $k\rightarrow k_{\mathrm{c%
}}$. Therefore, the system around EP should exhibit some peculiar critical
dynamics. The dynamics of $\mathcal{H}_{k_{\mathrm{c}}}$ alone cannot induce
any macroscopic phenomenon, while a set of $\mathcal{H}_{k}$\ near EP\ may
result in many-particle effect.

\section{P-wave pair generation}

\label{p-wave pair generation}

In this section, we investigate the critical dynamical behavior by applying
the obtained $U(t)$\ on a simple initial state. We start with the time
evolution of the vacuum state of operators $c_{\pm k}$\ as an initial state,
i.e.,%
\begin{equation}
\left\vert \psi _{k}(0)\right\rangle =\left\vert 0\right\rangle
_{k}\left\vert 0\right\rangle _{-k}.  \label{initial state}
\end{equation}%
In the case of $k\neq k_{\mathrm{c}}$, we have%
\begin{eqnarray}
\left\vert \psi _{k}(t)\right\rangle &=&U_{k}(t)\left\vert \psi
_{k}(0)\right\rangle \\
&=&\left[ -2i\sin \left( \varepsilon _{k}t\right) \sin ^{2}\theta _{k}+\exp
\left( i\varepsilon _{k}t\right) \right] \left\vert 0\right\rangle
_{k}\left\vert 0\right\rangle _{-k}  \notag \\
&&-\sin \left( 2\theta _{k}\right) \sin \left( \varepsilon _{k}t\right)
\left\vert 1\right\rangle _{k}\left\vert 1\right\rangle _{-k},  \notag
\end{eqnarray}%
while for $k=k_{\mathrm{c}}$, we have%
\begin{equation}
\left\vert \psi _{k_{\mathrm{c}}}(t)\right\rangle =\left( 1+it\right)
\left\vert 0\right\rangle _{k_{\mathrm{c}}}\left\vert 0\right\rangle _{-k_{%
\mathrm{c}}}-it\left\vert 1\right\rangle _{k_{\mathrm{c}}}\left\vert
1\right\rangle _{-k_{\mathrm{c}}}.  \label{kc_fs}
\end{equation}%
Accordingly, considering a vacuum state of all fermions (empty state) as an
initial state%
\begin{equation}
\left\vert \Psi (0)\right\rangle =\prod_{k>0}\left\vert \psi
_{k}(0)\right\rangle =\prod_{k>0}\left\vert 0\right\rangle _{k}\left\vert
0\right\rangle _{-k},
\end{equation}%
we have%
\begin{equation}
\left\vert \Psi (t)\right\rangle =\prod_{k>0}U_{k}(t)\left\vert \psi
_{k}(0)\right\rangle .
\end{equation}%
It is expected $p$-wave pairs are generated from the empty state.\ We are
interested in the normalized population of $p$-wave pair%
\begin{equation}
N(t)=\frac{\left\langle \Psi (t)\right\vert \widehat{N}\left\vert \Psi
(t)\right\rangle }{\left\langle \Psi (t)\right\vert \Psi (t)\rangle }%
=\sum_{k>0}\frac{\left\langle \Psi (t)\right\vert \widehat{N}_{k}\left\vert
\Psi (t)\right\rangle }{\left\langle \Psi (t)\right\vert \Psi (t)\rangle },
\end{equation}%
where the total $p$-wave pair number operator is
\begin{equation}
\widehat{N}=\sum_{k>0}\widehat{N}_{k}=\sum_{k>0}n_{k}n_{-k}.
\end{equation}%
Then $N(t)$ can be evaluated from $N_{k}(t)$
\begin{equation}
N(t)=\sum_{k>0}N_{k}(t)=\sum_{k>0}\frac{\left\langle \psi _{k}(t)\right\vert
\widehat{N}_{k}\left\vert \psi _{k}(t)\right\rangle }{\left\langle \psi
_{k}(t)\right\vert \psi _{k}(t)\rangle },
\end{equation}%
and the distribution of $N_{k}(t)$\ determines the property of the
non-equilibrium state.

For the case of $k\neq k_{\mathrm{c}}$, we have%
\begin{equation}
\left\langle \Psi _{k}(t)\right\vert \Psi _{k}(t)\rangle =2\left\vert \sin
2\theta _{k}\right\vert ^{2}\sin ^{2}\left( \varepsilon _{k}t\right) +1,
\end{equation}%
and%
\begin{equation}
N_{k}(t)=\frac{\left[ \Delta \sin k\sin \left( t\varepsilon _{k}\right) %
\right] ^{2}}{\left( \varepsilon _{k}/2\right) ^{2}+2\left[ \Delta \sin
k\sin \left( t\varepsilon _{k}\right) \right] ^{2}},  \label{Nkt}
\end{equation}%
which is a periodic function of time with period $T_{k}=\pi /\varepsilon
_{k} $\textbf{.} We note that we have $\varepsilon _{k}\approx 0$\ in the
vicinity of $k\approx k_{\mathrm{c}}$, and the period become very long. It
indicates that we always have $N_{k}(t)\approx 1/2$\ except for some short
intervals.

For the case of $k=k_{\mathrm{c}}$, Eq. (\ref{kc_fs}) shows that the
normalized pair number is
\begin{equation}
N_{k_{\mathrm{c}}}(t)=\frac{t^{2}}{1+2t^{2}},
\end{equation}%
which obeys \textrm{lim}$_{t\rightarrow \infty }N_{k_{\mathrm{c}}}(t)=1/2$,
which accords with the case with $k\neq k_{\mathrm{c}}$\ but infinite long
period. In order to demonstrate the property of the evolved state, we define
the average normalized pair number distribution%
\begin{equation}
\overline{N}_{k}=\frac{1}{T_{k}}\int_{0}^{T_{k}}N_{k}(t)\mathrm{d}t,
\label{aNk}
\end{equation}%
and the\ total average normalized pair number%
\begin{equation}
\overline{N}\left( t\right) =\frac{1}{\pi }\int_{0}^{\pi }N_{k}(t)\mathrm{d}%
k.  \label{aNt}
\end{equation}%
We plot quantities $\overline{N}_{k}$, $\overline{N}(t)$, and $N_{k}(t)$\
for a concrete cases in Fig. \ref{fig2}.\ It indicates that the majority of
modes become quasi stable after a period of time. Accordingly, the evolved
many-body state $\left\vert \Psi (t)\right\rangle $\ should exhibit\ as a
macroscopic equilibrium state. In the following section, we will investigate
the possible property of such a state.

\section{Dynamical generation of superconducting state}

\label{superconducting state}
\begin{figure}[tbp]
\includegraphics[ bb=24 0 376 350, width=0.3\textwidth, clip]{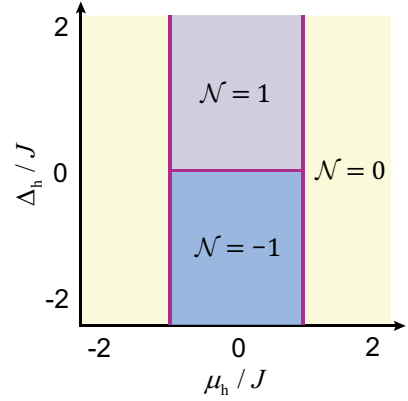}
\caption{Phase diagram of Hermitian Kitaev model on the parameter $\mu_h -\Delta_h $\ plane. The skyblue, yellow and purple regions correspond to the
winding number $-1$, $0$ and $1$ respectively. Darkmagenta lines indicate
the phase transition lines.}
\label{fig3}
\end{figure}

In this section, as an application of above
result, we investigate the possibility of dynamical generation of
superconducting state via a non-Hermitian Kitaev model. The scheme is that
taking the empty state $\sum_{k>0}\left\vert 0\right\rangle _{k}\left\vert
0\right\rangle _{-k}$ as an initial state, the final state, which approches
to the ground state of a Hermitian Kitaev Hamiltonian $H$, is acheived by a
driven non-Hermitian Kitaev Hamiltonian $\mathcal{H}$ at EP. Before
proceeding, we briefly review the properties of a Hermitian Kitaev model
with the Hamiltonian

\begin{eqnarray}
H &=&\sum\limits_{j=1}^{N}[-Jc_{j}^{\dag }c_{j+1}+\mathrm{H.c.}-i\Delta _{%
\mathrm{h}}c_{j}^{\dag }c_{j+1}^{\dag } \\
&&+i\Delta _{\mathrm{h}}c_{j+1}c_{j}+\mu _{\mathrm{h}}\left( 2n_{j}-1\right)
].  \notag
\end{eqnarray}%
It has been shown to have topologically non-trivial (trivial) ground state,
when $\left\vert \mu _{\mathrm{h}}\right\vert <\left\vert J\right\vert $ ($%
\left\vert \mu _{\mathrm{h}}\right\vert >\left\vert J\right\vert $) in Ref.
\cite{Kitaev}. The phase diagram is plotted in Fig. \ref{fig3}, with H-shape
boundary separating topologically non-trivial and trivial phases,
characterized by winding number $\mathcal{N}$. By the similar procedure as
above, we have%
\begin{eqnarray}
H &=&\sum_{\pi >k>0}H_{k}, \\
H_{k} &=&2\left(
\begin{array}{cc}
c_{k}^{\dag } & c_{-k}%
\end{array}%
\right) \left(
\begin{array}{cc}
\mu _{\mathrm{h}}-J\cos k & \Delta _{\mathrm{h}}\sin k \\
\Delta _{\mathrm{h}}\sin k & J\cos k-\mu _{\mathrm{h}}%
\end{array}%
\right) \left(
\begin{array}{c}
c_{k} \\
c_{-k}^{\dag }%
\end{array}%
\right) ,  \notag \\
&&
\end{eqnarray}%
where the Hamiltonian $H_{k}$\ in each invariant subspace satisfies the
commutation relation%
\begin{equation}
\left[ H_{k},H_{k^{\prime }}\right] =0.
\end{equation}%
For a given $k$, the Hamiltonian $H_{k}$\ in the basis ($\left\vert
0\right\rangle _{k}\left\vert 0\right\rangle _{-k}$, $\left\vert
1\right\rangle _{k}\left\vert 1\right\rangle _{-k}$, $\left\vert
1\right\rangle _{k}\left\vert 0\right\rangle _{-k}$, $\left\vert
0\right\rangle _{k}\left\vert 1\right\rangle _{-k}$) is expressed as $%
4\times 4$ matrix%
\begin{equation}
h_{k}=2\left(
\begin{array}{cccc}
J\cos k-\mu _{\mathrm{h}} & \Delta _{\mathrm{h}}\sin k & 0 & 0 \\
\Delta _{\mathrm{h}}\sin k & \mu _{\mathrm{h}}-J\cos k & 0 & 0 \\
0 & 0 & 0 & 0 \\
0 & 0 & 0 & 0%
\end{array}%
\right) .
\end{equation}

\begin{figure*}[tbp]
\includegraphics[ bb=23 73 695 373, width=0.95\textwidth, clip]{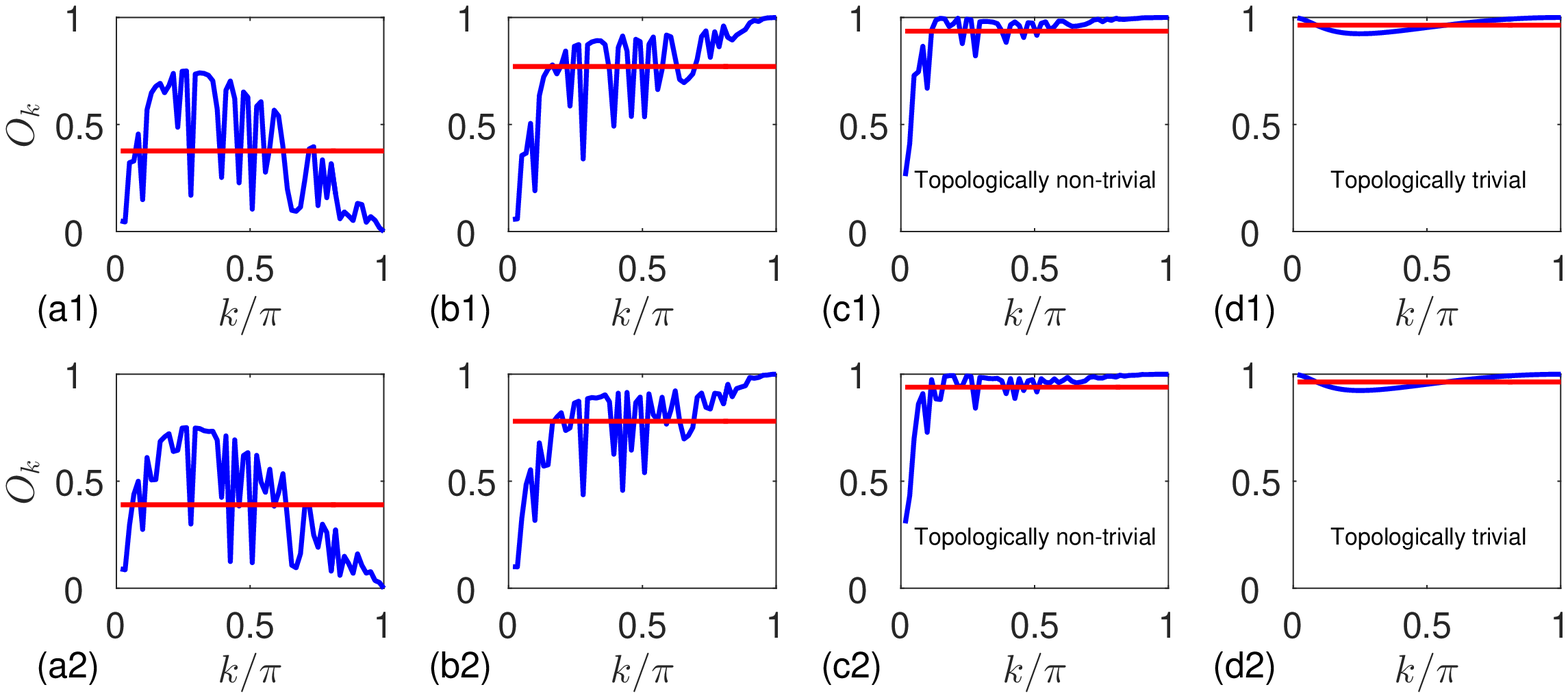} %
\includegraphics[ bb=23 73 695 220, width=0.95\textwidth, clip]{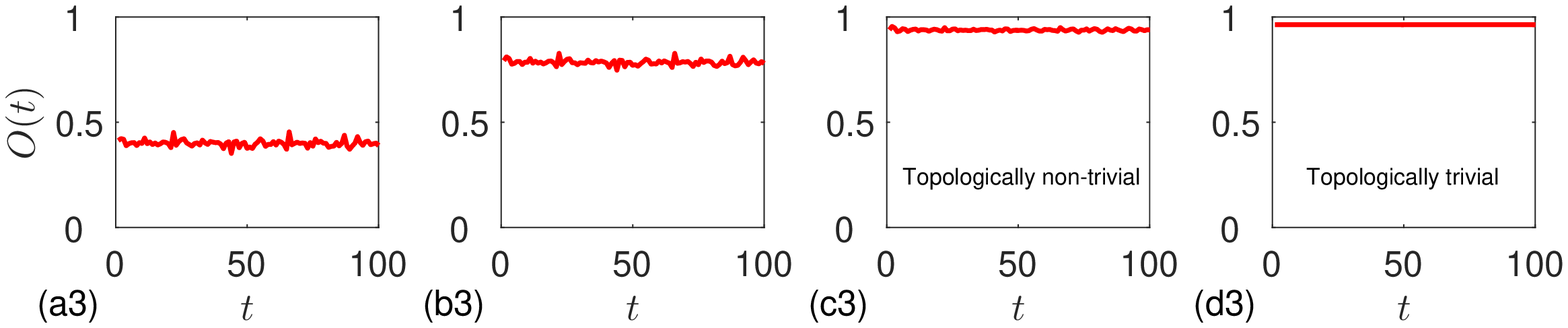}
\caption{Numerical simulations of $O_{k}$ defined in Eq. (\protect\ref{Okt}%
),\ and $O\left( t\right) $ defined in Eq. (\protect\ref{Ot}).\ The four
panels in first row and the second row are the plots of $O_{k}$\ at time $%
t=50J^{-1}$ and $100J^{-1}$, respectively. The red lines represent the
corresponding $O$, which are the values (a1) $0.376$, (a2) $0.390$, (b1) $%
0.771$, (b2) $0.780$, (c1) $0.936$, (c2) $0.939$, (d1) $0.964$, (d2) $0.964$%
. The four panels in third row are the plots of $O(t)$\ with the same
parameters in two above rows. The parameters are $N=61$, $J=1$, $\Delta
=\Delta _{\mathrm{h}}=1$, and $\protect\mu =\protect\sqrt{J^{2}+\Delta ^{2}}$%
. Each column of the graph has the same set of parameters, i.e., (a1-a3) $%
\protect\mu _{\mathrm{h}}=-5$, (b1-b3) $\protect\mu _{\mathrm{h}}=-0.5$, and
(c1-c3) $\protect\mu _{\mathrm{h}}=0.9$, (d1-d3) $\protect\mu _{\mathrm{h}}=%
\protect\mu $. The parameters of Hermitian Kitaev model in (c1-c3) and (d1-d3)
supports topologically non-trivial and trivial superconducting ground
states, respectively.}
\label{fig4}
\end{figure*}

\bigskip The eigenstates $\left\vert \varphi _{k\mathrm{\lambda }}^{\pm
}\right\rangle $\ ($\mathrm{\lambda }=\mathrm{e,o}$ denotes the even/odd
parity of the particle number) are

\begin{eqnarray}
\left\vert \varphi _{k\mathrm{e}}^{\pm }\right\rangle  &=&\frac{1}{\sqrt{%
\Omega _{\mathrm{h}}^{\pm }}}\left( \left\vert 0\right\rangle _{k}\left\vert
0\right\rangle _{-k}+b_{k}^{\pm }\left\vert 1\right\rangle _{k}\left\vert
1\right\rangle _{-k}\right) , \\
\left\vert \varphi _{k\mathrm{o}}^{+}\right\rangle  &=&\left\vert
1\right\rangle _{k}\left\vert 0\right\rangle _{-k},\left\vert \varphi _{k%
\mathrm{o}}^{-}\right\rangle =\left\vert 0\right\rangle _{k}\left\vert
1\right\rangle _{-k},
\end{eqnarray}%
where $\Omega _{\mathrm{h}}^{\pm }=1+\left\vert b_{k}^{\pm }\right\vert ^{2}$
is the normalization coefficient in the context of Dirac inner product with%
\begin{equation}
b_{k}^{\pm }=\frac{\Delta _{\mathrm{h}}\sin k}{J\cos k-\mu _{\mathrm{h}}\pm
\epsilon _{k\mathrm{e}}^{\pm }/2},
\end{equation}%
and corresponding energies are

\begin{equation}
\left\{
\begin{array}{c}
\epsilon _{k\mathrm{e}}^{\pm }=\pm 2\sqrt{\left( \mu _{\mathrm{h}}-J\cos
k\right) ^{2}+\Delta _{\mathrm{h}}^{2}\sin ^{2}k}, \\
\epsilon _{k\mathrm{o}}^{\pm }=0.%
\end{array}%
\right.
\end{equation}%
Accordingly, the groundstate wave function can be expressed as%
\begin{equation}
\left\vert \text{\textrm{G}}\right\rangle =\prod_{\pi >k>0}\left\vert
\varphi _{k\mathrm{e}}^{-}\right\rangle .
\end{equation}%
We note that for a topological non-trivial ground state, we have%
\begin{equation}
\lim_{k\rightarrow 0}\left\vert \varphi _{k\mathrm{e}}^{-}\right\rangle
=\left\vert 1\right\rangle _{k}\left\vert 1\right\rangle
_{-k},\lim_{k\rightarrow \pi }\left\vert \varphi _{k\mathrm{e}%
}^{-}\right\rangle =\left\vert 0\right\rangle _{k}\left\vert 0\right\rangle
_{-k},
\end{equation}%
while\textbf{\ }%
\begin{equation}
\lim_{k\rightarrow 0}\left\vert \varphi _{k\mathrm{e}}^{-}\right\rangle
=\left\vert 0\right\rangle _{k}\left\vert 0\right\rangle
_{-k},\lim_{k\rightarrow \pi }\left\vert \varphi _{k\mathrm{e}%
}^{-}\right\rangle =\left\vert 0\right\rangle _{k}\left\vert 0\right\rangle
_{-k},
\end{equation}%
for a topological trivial ground state.\ On the other hand, for the
non-Hermitian system, we know that there is a stable final state $%
\lim_{t\rightarrow \infty }\left\vert \psi _{k_{\mathrm{c}}}(t)\right\rangle
\propto (\left\vert 0\right\rangle _{k_{\mathrm{c}}}\left\vert
0\right\rangle _{-k_{\mathrm{c}}}-\left\vert 1\right\rangle _{k_{\mathrm{c}%
}}\left\vert 1\right\rangle _{-k_{\mathrm{c}}})$, according to Eq. (\ref%
{kc_fs}). If we take a matching set of parameters, the stable final state
can be an eigenmode of $\left\vert \text{\textrm{G}}\right\rangle $ , i.e., $%
\left\vert \psi _{k_{\mathrm{c}}}(t)\right\rangle =\left\vert \varphi _{k_{%
\mathrm{c}}\mathrm{e}}^{-}\right\rangle $\ after normalization.\textbf{\ }It
is probably to obtain a state dynamically under the Hamiltonian $\mathcal{H}$%
, which is similar to a ground state of $H$. To characterize how close of an
evolved state to a superconducting state we introduce a quantity%
\begin{equation}
O(t)=\frac{1}{N}\sum_{k}O_{k}(t),  \label{Ot}
\end{equation}%
where%
\begin{equation}
O_{k}(t)=\left\langle \varphi _{k\mathrm{e}}^{-}\right. \left\vert \psi
_{k}(t)\right\rangle ,  \label{Okt}
\end{equation}%
is the overlap of a specific topological superconducting mode $\left\vert
\psi _{k\mathrm{e}}^{-}\right\rangle $\ and a dynamically generated state $%
\left\vert \psi _{k}(t)\right\rangle $ via the non-Hermitian system.

We compute the quantity $O(t)$\ for various sets of parameters $\left(
J,\Delta ,\mu \right) $\ and $\left( J,\Delta _{\mathrm{h}},\mu _{\mathrm{h}%
}\right) $\ to search optimal cases with large $O(t)$. We find that there
are many cases with large $O(t)$. Here we take four typical cases to
demonstrate our results. We plot $O(t)$\ and $O_{k}$\ at certain instants in
Fig. \ref{fig4}, which show that $O(t)$\ oscillates with a very small
amplitude. It also indicates that through such a dynamical method, a
quasi-superconducting state involving topological trivial and non-trivial
can be generated from a simple initial state.

\section{Summary}

\label{sec_summary}

In summary, we have studied the non-Hermitian extension of Kitaev chain by
considering imaginary $p$-wave pairing amplitudes. Based on the analysis of
the exact solution we find that exceptional line is hyperbolic, which
separates two regions with real and complex\ Bogoliubov-de-gens spectra,
associated with $\mathcal{PT}$-symmetry breaking. The EPs are movable in $k$
space as the parameters vary along the exceptional line. The non-Hermiticity
around EP supports resonant generation of $p$-wave Cooper pair state via the
critical dynamic process. A specific pair state $(1+c_{k}^{\dagger
}c_{-k}^{\dagger })\left\vert 0\right\rangle $\ with selecting momentum $k$
can be generated from the vacuum state $\left\vert 0\right\rangle $ of
fermions and be frozen forever. The remarkable result obtained by analytical
approaches and numerical simulations are that the dynamically generated
state via the non-Hermitian system is very close to a specific
superconducting ground state, which can be topologically non-trivial or not.
This finding provides alternative way to generate a superconducting state
via critical dynamic process rather than cooling down the temperature.

\section*{Acknowledgement}

We acknowledge the support of NSFC (Grants No. 11874225).

\end{document}